\documentclass[secnumarabic, graphics,floatfix, nofootinbib,tightenlines,nobibnotes, aps, prl, 12pt]{revtex4-1}
\usepackage{graphicx}
\usepackage[english]{babel}
\usepackage[utf8]{inputenc}
\usepackage{mathrsfs}
\usepackage{scalerel,amsmath,amssymb}
\usepackage{amsfonts}
\usepackage{multirow}
\usepackage[section]{placeins}

\newcommand{\bq}{\begin{equation}}
\newcommand{\eq}{\end{equation}}
\newcommand{\bqn}{\begin{eqnarray}}
\newcommand{\eqn}{\end{eqnarray}}

\newcommand{\lb}{\label}

\def\msquare{\mathord{\scalerel*{\Box}{gX}}}

\begin{document}

\title{On quasinormal modes for the Vaidya metric in asymptotically anti-de Sitter spacetime}

\author{Kai Lin$^{1,2}$}
\author{Yunqi Liu$^{3}$}
\author{Wei-Liang Qian$^{2,4,3}$}
\author{Bin Wang$^{3,5}$}
\author{Elcio Abdalla$^{6}$}

\affiliation{$^{1}$ Hubei Subsurface Multi-scale Imaging Key Laboratory, Institute of Geophysics and Geomatics, China University of Geosciences, 430074, Wuhan, Hubei, China}
\affiliation{$^{2}$ Escola de Engenharia de Lorena, Universidade de S\~ao Paulo, 12602-810, Lorena, SP, Brazil}
\affiliation{$^{3}$ Center for Gravitation and Cosmology, School of Physical Science and Technology, Yangzhou University, 225002, Yangzhou, Jiangsu, China}
\affiliation{$^{4}$ Faculdade de Engenharia de Guaratinguet\'a, Universidade Estadual Paulista, 12516-410, Guaratinguet\'a, SP, Brazil}
\affiliation{$^{5}$ School of Aeronautics and Astronautics, Shanghai Jiao Tong University, 200240, Shanghai, China}
\affiliation{$^{6}$ Instituto de F\'isica, Universidade de S\~ao Paulo, C.P. 66318, 05315-970, S\~ao Paulo, SP, Brazil}

\date{Aug. 31, 2019}

\begin{abstract}
In this work, we present a numerical scheme to study the quasinormal modes of the time-dependent Vaidya black hole metric in asymptotically anti-de Sitter spacetime.
The proposed algorithm is primarily based on a generalized matrix method for quasinormal modes.
The main feature of the present approach is that the quasinormal frequency, as a function of time, is obtained by a generalized secular equation and therefore a satisfactory degree of precision is achieved. 
The implications of the results are discussed.
\end{abstract}

\maketitle

\section{I. Introduction}

It is understood that the quasinormal modes are eigenmodes of a system subject to internal dissipation or energy radiation.
In terms of the temporal evolution of small perturbations, the amplitude of the oscillation decays in time.
Owing to the damping characteristic, the frequency of a quasinormal mode is complex, where its imaginary part is positive for stable configurations and suppresses the oscillations.
In general relativity, small perturbations of a black hole, in terms of external matter field or metric perturbations, generally produce quasinormal modes~\cite{agr-qnm-review-01,agr-qnm-review-02,agr-qnm-review-03}. 
In this case, damping takes place not by internal friction, but through radiation of energy towards infinity or into the black hole.
In particular, the recent development of the holographic principle regarding the anti-de Sitter/conformal field theory (AdS/CFT) correspondence~\cite{agr-qnm-holography-review-01} has further promoted extensive studies.
As the AdS/CFT correspondence is an essential tool for exploring the strongly coupled systems, it can be employed to investigate the fundamental properties of the system.
In principle, various transport coefficients of the dual system can be extracted, such as the viscosity, conductivity, and diffusion constants. 
Moreover, the first detection of gravitational waves~\cite{agr-LIGO-01} has driven the relevant studies into a direction directly associated with precise measurements.

From a mathematical point of view, analyzing quasinormal modes involves the solution of non-Hermitian eigenvalues regarding a system of coupled linear ordinary differential equations with appropriate boundary conditions.
Aside from a few analytic solutions, in order to evaluate the quasinormal frequencies, one usually has to resort to numerical methods~\cite{agr-qnm-review-04}.
Many numerical techniques have been proposed.
Among others are the WKB method~\cite{agr-qnm-WKB-01,agr-qnm-WKB-02,agr-qnm-WKB-03}, the continued fraction method~\cite{agr-qnm-continued-fraction-01,agr-qnm-continued-fraction-02}, the Poshl-Teller potential approximation~\cite{agr-qnm-Poshl-Teller-01}, the Horowitz and Hubeny (HH) method for AdS black holes~\cite{agr-qnm-HH-01}, the matrix method~\cite{agr-qnm-lq-matrix-02,agr-qnm-lq-matrix-03}.
For the study of the temporal evolution of the small perturbations, the finite difference method can be employed~\cite{agr-qnm-finite-difference-01,agr-qnm-finite-difference-02}.

In general, black holes are dynamic rather than static objects.
Primordial black holes, which possess the size $\sim H^{-1}$, are intrinsically dynamical.
The first observation of gravitational waves matches the predictions~\cite{agr-merger-01,agr-merger-02,agr-merger-03} for a gravitational wave emanating from the merger of a pair of black holes.
Also, mass accretion cause the mass of the astrophysical black hole to evolve in time (in principle there is also Hawking radiation, but it is negligible for astrophysical black holes).
In this context, the analysis of quasinormal modes for time-dependent situations is of particular interest.
The Vaidya metric provides an asymptotically flat and spherically symmetric solution of the Einstein equations describing the spacetime outside of a star, which accretes or radiates pressureless null dust.
The metric has been employed as an essential tool to explore dynamical processes, such as black-hole evaporation including Hawking radiation~\cite{agr-metric-vaidya-app-01,agr-metric-vaidya-app-02,agr-hawking-radiation-03}.
Besides, it has been used as one of the possibilities~\cite{agr-qnm-time-dependent-03,agr-qnm-time-dependent-04,agr-qnm-time-dependent-05} to investigate the time-dependent black hole quasinormal modes~\cite{agr-qnm-time-dependent-01,agr-qnm-time-dependent-02}.
Most studies concerning the quasinormal modes have been carried out by using the finite difference method, where the boundary of the problem is transformed to infinity, and therefore free boundary condition has been employed.
The corresponding quasinormal frequencies are subsequently extracted numerically by using $\chi^2$ fitting. 
Instead, for the present study, the boundary condition is treated explicitly, particularly for that at the apparent horizon.
Moreover, we introduce a numerical scheme to obtain the quasinormal frequencies as well as the corresponding wave function by solving a matrix equation.
As a result, the proposed approach provides reliable precision which can be easily generalized to other scenarios of dynamic black holes. 

The primary purpose of the present study is to present the numerical scheme and use it to investigate the quasinormal modes of time-dependent backgrounds associated with the Vaidya metric in asymptotically AdS spacetime.
The paper is organized as follows.
In the next section, we derive the master equation for scalar perturbations for the time-dependent case and compare it to the corresponding static situation.
Then, we discretize the spatial and time coordinates and reformulate the partial differential equation in terms of a matrix equation.
The numerical scheme is thereby presented.
In section III, the numerical results are obtained and discussed with particular emphasis on the nonstationary effects.
Further discussions on the implications of the present approach, as well as concluding remarks, are given in section IV.

\section{II. Quasinormal frequency for the Vaidya black hole} \label{Sect2}

In terms of the Eddington coordinates, the metric of Vaidya AdS spacetime reads~\cite{agr-metric-vaidya-01,agr-metric-vaidya-02,agr-metric-vaidya-03,Zhao:1992ad,Zhao:1994hm,Wang:1998qx}
\bqn
ds^2=-f(v,r)dv^2+2cdrdv+r^2(d\theta^2+\sin^2\theta d\varphi^2) ,\lb{leDmetric}
\eqn
with
\bqn
f(v,r)= 1-\frac{2M(v)}{r}-\frac{\Lambda}{3}r^2 ,\lb{Vmetric}
\eqn
where $\Lambda<0$, and without any loss of generality, we choose $\Lambda=-3$ in the following calculations. On the other hand, $c=\pm 1$.
To be specific, $c=1$ corresponds to the case of ingoing flow and $M(v)$ is a monotonically increasing function of the advanced time, while $c=-1$ corresponds to the case of outgoing flow and $M(v)$ is a monotonically decreasing function of the retarded time.
The master equation for small perturbations of a massive scalar field is governed by the Klein-Gordon equation which reads
\bqn
(\msquare + m_\mu^2)  \Psi = 0 ,
\eqn
or,
\bqn
\frac{1}{\sqrt{-g}}{\partial_\mu}\left(g^{\mu\nu}\sqrt{-g}{\partial_\nu\Psi}\right)-m_\mu^2\Psi=0 .
\eqn

One proceeds by using the method of separation of variables which assumes
\bqn
\Psi = \frac{\Phi(r,v)}{r}Y(\theta, \varphi) ,\lb{sepv}
\eqn
where the radial part of the wave function $\Phi$ is assumed to be time dependent.
The angular part of the wave function $Y(\theta, \varphi)=\Theta(\theta)\exp\left[im\varphi\right]$ are simply the spherical harmonics satisfying
\bqn
\sin\theta\frac{d}{d\theta}\left[\sin\theta\frac{d\Theta(\theta)}{d\theta}\right]+\ell(\ell+1)\sin^2\theta\Theta(\theta)-m^2\Theta(\theta) = 0 ,
\eqn
where $\ell$ and $m$ are the azimuthal and magnetic quantum numbers respectively.
By substituting Eq(\ref{sepv}) as well as the metric, $\Phi$ is found to satisfy the equation
\bqn
-\left[m_\mu^2+\frac{\ell+\ell^2+rf'}{r^2}\right]\Phi+f'\Phi'+2c{\dot \Phi}'+f\Phi'' = 0 .\lb{master0}
\eqn
where `` $ ' $ " indicates partial derivative with respect to $r$ and ``$\cdot$" indicates partial derivative with respect to $v$.
To investigate the boundary condition, let us consider the case $c=1$. 
One notices that the above equation can be rewritten as
\bqn
f (f\Phi ' ) '+ 2f {\dot \Phi}' = \frac{1}{r^2}f\Phi (\ell+\ell^2 + m_\mu^2 r^2+r f') . \lb{master1}
\eqn
The r.h.s. of the above equation vanishes as one approaches either the apparent horizon or infinity.
In other words, near the horizon and infinity, the master equation reads
\bqn
f (f\Phi ' ) '+ 2f {\dot \Phi}' = 0 .
\eqn
The general solution of the above equation is $C_1\Phi_1+C_2\Phi_2$, where $C_1, C_2$ are two constants, $\Phi_1=e^{-i\omega(v)v}$ and $\Phi_2$ satisfies $f\Phi'_2+2{\dot \Phi_2} = 0$.
At the horizon, only the ingoing waves are physically permitted, and therefore only $\Phi_1$ is relevant.
At infinity, on the other hand, the wave function approaches zero for asymptotically AdS spacetime.
Accordingly, it is reasonable to assume that the solution of Eq.~(\ref{master0}) possesses the form
\bqn
\Phi(r,v) = e^{-i\omega(v)v}R(r,v) .\lb{sepv2}
\eqn
where the quasinormal frequency $\omega=\omega(v)$ is expected to be time dependent.

The equation of $R(r,v)$ can be obtained straightforwardly from Eq.~(\ref{master0}), which reads
\bqn
-\left[m_\mu^2+\frac{\ell+\ell^2+rf'}{r^2}\right]R+(-2i\omega-2iv{\dot\omega}+f')R'+2{\dot R}'+fR'' = 0 . \lb{masterEqVaidya}
\eqn
The corresponding boundary conditions are~\cite{agr-qnm-lq-matrix-04}
\bq
\lb{qnmbc}
R\sim 
\left\{\begin{array}{cc}
0      &  r\rightarrow \infty   \cr\\
C_1 &  r\rightarrow r_h
\end{array}\right. , 
\eq
where $C_1$ is time independent.

It is not difficult to show that Eq.~(\ref{masterEqVaidya}) falls back to that of scalar perturbation in static Schwarzschild AdS black hole spacetimes, namely,
\bqn
-\left[m_\mu^2+\frac{\ell+\ell^2+rf'}{r^2}\right]R+(-2i\omega+f')R'+fR'' = 0 , \lb{masterEqStaticAdS}
\eqn
by eliminating all the terms involving time derivative.
Eq.~(\ref{masterEqStaticAdS}) and its solution will be addressed below while we discuss the numerical results in the following section.

Before presenting the numerical scheme to solve the master equation, we comment further about its boundary conditions.
First of all, we note that, after canceling the factor $e^{-i\omega(v)v}$, the resulting boundary condition for $R(r,v)$ does not depend on $v$, which turns out to be quite useful to facilitate the present algorithm.
For a dynamical black hole metric, the apparent and event horizons usually do not coincide.
The apparent horizon is defined as the outer component of the intersection of the trapped region and a spacelike surface~\cite{book-blackhole-physics-Frolov}.
At a given instant, it is a surface that plays the role of the boundary separating the light rays that are directed outwards and moving outwards, and those headed outward but moving inward.
The choice of the boundary condition for quasinormal modes is dictated by the condition that only ingoing waves are physically permitted, associated with the fact that classical horizons do not emit radiation~\cite{agr-qnm-01,agr-qnm-holography-review-01}.
In other words, out of two local solutions near the boundary, which typically represent the incoming as well as outgoing waves, one only chooses the incoming waves.
This choice leads to a profound consequence for the master equation.
To be more specific, the particular choice of the boundary condition implies that the corresponding boundary value problem is non-Hermitian, and subsequently, the associated eigenfrequencies become complex~\cite{agr-qnm-holography-review-01}.
In the case of the Vaidya black hole metric, following Refs.\cite{agr-metric-vaidya-04,agr-metric-vaidya-05}, the location of the apparent horizon can be determined by
\bqn
1-\frac{2M(v)}{r_h}+r_h^2 = 0 . \lb{eqAH}
\eqn
Eq.~(\ref{eqAH}) implies that the apparent horizon is moving outward if the black hole mass increases in time.
On the other hand, the event horizon $r_{\mathrm{EH}}$ is defined by the boundary of the region of spacetime from which no causal signal can escape to future null infinity ${\mathscr I}^+$.
As shown for particular parameters in the Vaidya metric~\cite{agr-qnm-time-dependent-04}, the apparent horizon mostly resides inside the event horizon.
Moreover, owing to the physical characteristic of the event horizon, for the case of a dynamic black hole metric, a matter flow directed outwards may actually traverse the event horizon.
On the contrary, the apparent horizon serves as a one-way membrane which prohibits even the outgoing light rays from traveling across it.
Concerning the context of quasinormal modes, where the matter flow is represented by the probability flow of the wave function, it is reasonable to introduce the boundary condition of the master equation, Eq.~(\ref{masterEqVaidya}), at the apparent horizon instead of the event horizon.
Therefore, one requires that the wave function must be ingoing at the apparent horizon $r_h$ as shown above in Eq.~(\ref{qnmbc}).

In order to solve Eq.~(\ref{masterEqVaidya}) with the boundary condition Eq.~(\ref{qnmbc}) defined at the apparent horizon Eq.~(\ref{eqAH}), we resort to a generalized version of the matrix method proposed recently~\cite{agr-qnm-lq-matrix-01,agr-qnm-lq-matrix-02,agr-qnm-lq-matrix-03,agr-qnm-lq-matrix-04}.
The time ($v$) derivative only involves the first order and is handled by the forward-difference formula.
The matrix method is employed to deal with spatial derivatives.
First, we transform the space coordinate $r$ into $x=r_h/r$ and rewrite the master equation in terms of $x$.
Since the resultant domain of the wave function, $0<x<1$, is finite, we discretize the wave function into $N+1$ grids.
According to the spirit of the matrix method, now any spatial derivative of the wave function has been transformed into a linear combination of the function values on the grids.
Therefore, by substituting these expressions into the master equation Eq.~(\ref{master0}) for each grid, at a given instant, the function values on grids and their temporal derivatives are related by an $(N+1)\times(N+1)$ matrix equation.
In other words, if the wave function $R(r,v)$ is known at a given instant $v=v_i$, one has $N+1$ equations which can be solved to obtain the wave function on the $N+1$ grids for the instant $v_{i+1}$, once the finite forward-difference discussed above is implemented for the first order time derivative.

However, if one carefully counts the number of variables and the number equations at hand, there is a subtlety.
The boundary conditions at the horizon and infinity eliminate two variables since according to Eq.~(\ref{qnmbc}) the function values at those two grids are time independent.
Regarding the two corresponding equations, the one at infinity is actually redundant and therefore is discarded.
As a result, one possesses $N$ equations from the discretized master equation and $N-1$ variables associated with all the grid points except two on the boundary.
In other words, we have one additional equation.
The latter can be conveniently utilized to determine the quasinormal frequency $\omega_{i+1}$ at the instant $v_{i+1}$, which completes our scheme.
We also note that the resulting equation for $\omega_{i+1}$ is merely an algebraic equation and can be solved easily by a numerical method.
 
\section{III. Numerical results} \label{Sect3}

\begin{figure}[ht]
\begin{minipage}{225pt}
\centerline{\includegraphics[width=225pt]{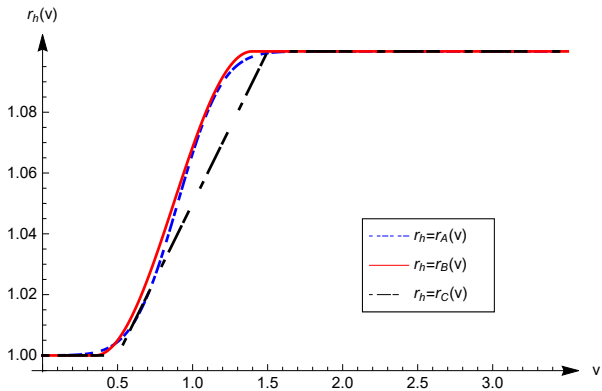}}
\end{minipage}
\caption{(Color online) The three different time-dependent functions for $r_h$ studied in the present work.
The calculations have been carried out by taking $r_1=1$ and $r_2=1.1$.
The specific forms of the functions, $r_A$, $r_B$, and $r_C$, shown in dashed blue, solid red, and dash-dotted black curves, are defined in Eq.~(\ref{massEx1}), Eq.~(\ref{massEx2}), and (\ref{massEx3}), respectively.}
\label{QNMf1}
\end{figure}

Now we proceed to implement the numerical scheme presented at the end of the last section to the metric Eq.~(\ref{Vmetric}), where we consider the following mass function in terms of the apparent horizon $r_h$ as a function of time, as shown by the dashed blue curve in Fig.~\ref{QNMf1}
\bqn
M(v)= \frac{r_h(v)^3 + r_h(v)}{2} ,\lb{eqAH2}
\eqn
with
\bqn
r_h(v) = r_A(v)\equiv r_1 + (r_2-r_1) \frac12\left[\mathrm{erf} \left(C(v-v_1)\right)+1\right]  ,\lb{massEx1}
\eqn
where $\mathrm{erf}$ is the error function, numerically, we adopt $v_1=0.9$ and $C=3$.
Here $r_h$ evolves smoothly from $r_1$ to $r_2$ for the interval $-\infty \le v < +\infty$.
Eq.~(\ref{massEx1}) implies that the black hole mass remains a constant for an infinitely long period and therefore it is essentially ``static" for $v < 0$ with an appropriately chosen $v_1$.
As a result, the solution of the quasinormal problem of a static black hole metric with $M_1=\frac{r_1^3 + r_1}{2}$ is utilized as the initial condition for the present dynamic case.
To be specific, Eq.~(\ref{masterEqStaticAdS}) is solved by employing the matrix method in its original form~\cite{agr-qnm-lq-matrix-01,agr-qnm-lq-matrix-02,agr-qnm-lq-matrix-03}, and its solution, $\omega$ and $R(r)$, is fed to the proposed scheme for solving Eq.~(\ref{masterEqVaidya}). 
From the instant $v=0$ onward, we employ the matrix method to interpolate the spatial derivatives and forward-difference formula for the time evolution.
For simplicity, the calculations are carried out for the perturbations of a massless sacalar field.

But before discussing the properties of the quasinormal modes of dynamical black holes, it is meaningful to show that the results regarding the physical system are manifestly convergent. 
In other words, the obtained numerical results should not be sensitive to small deviations of the chosen mass function.
This is achieved by carrying out the calculations also by two slightly different parameterizations, whose forms have been adopted in some previous studies~\cite{agr-qnm-time-dependent-02,agr-qnm-time-dependent-03,agr-qnm-time-dependent-04}.
Moreover, although the matrix method has shown to be up to par in various studies of quasinormal modes of static black holes, one should also warrant the precision of the numerical scheme for the case of dynamical black holes.
We relegate these studies to the Appendix of the paper.

\begin{figure*}[h]
\includegraphics[width=6cm]{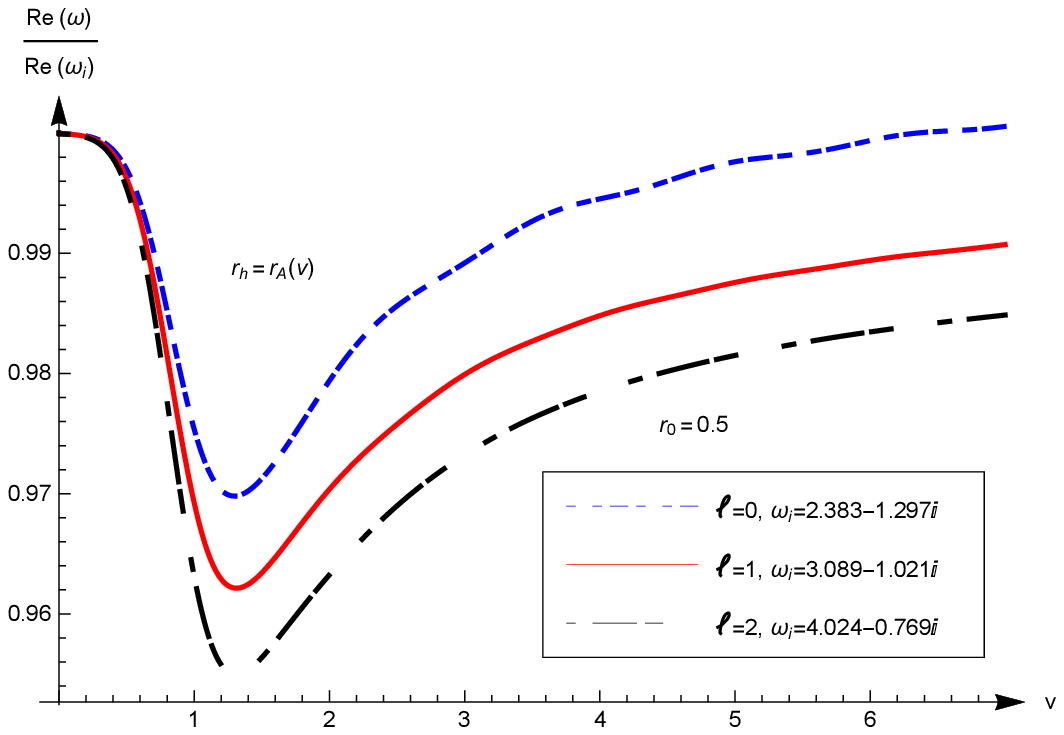}\includegraphics[width=6cm]{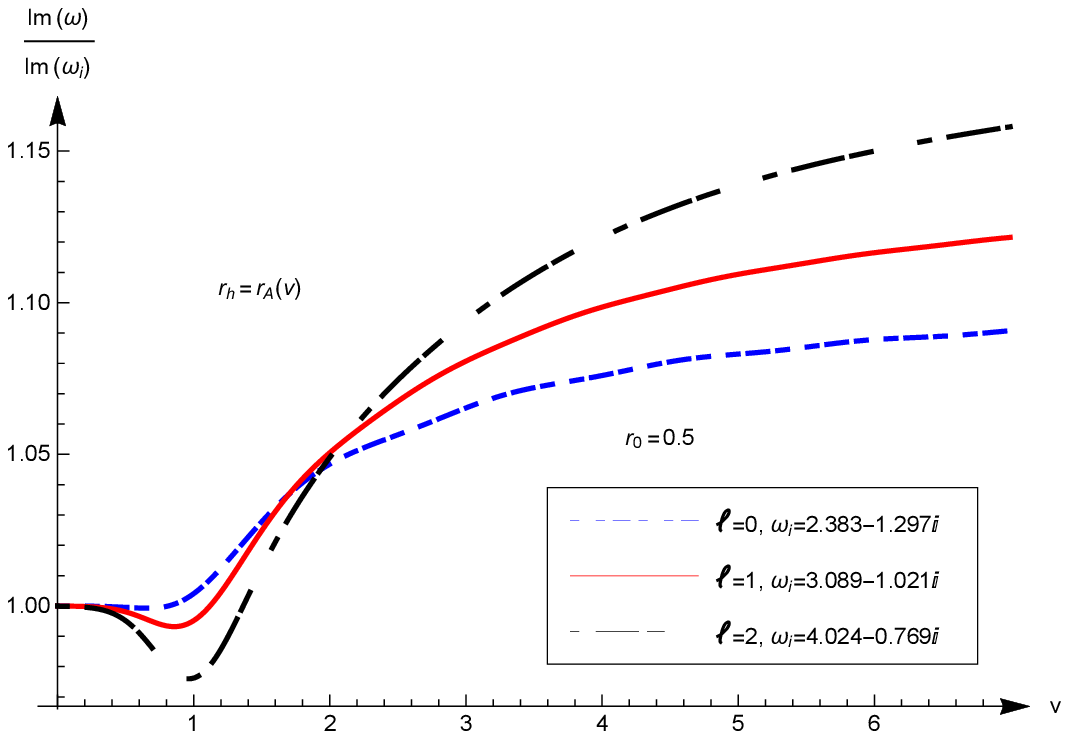}
\includegraphics[width=6cm]{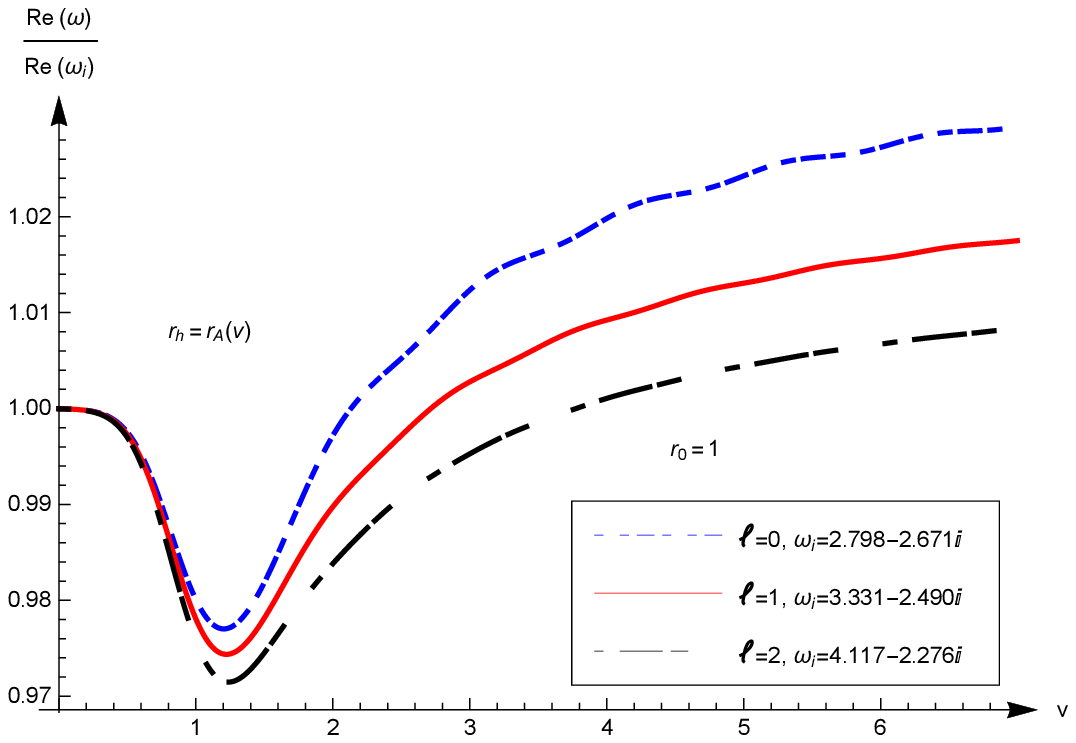}\includegraphics[width=6cm]{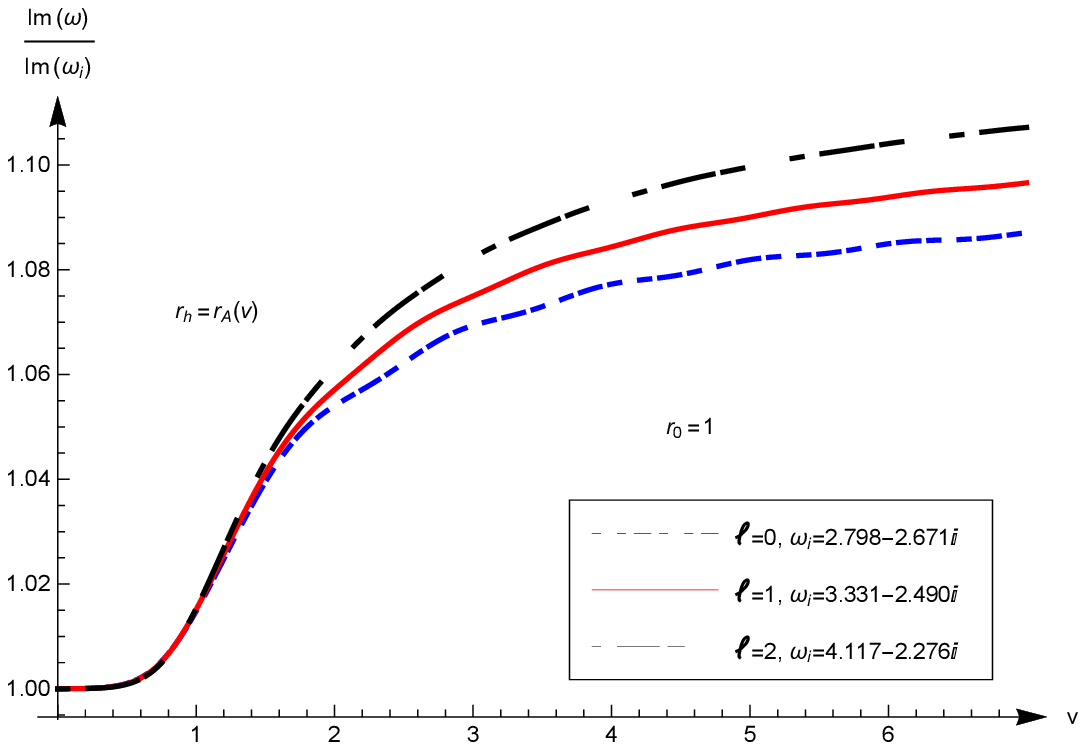}
\includegraphics[width=6cm]{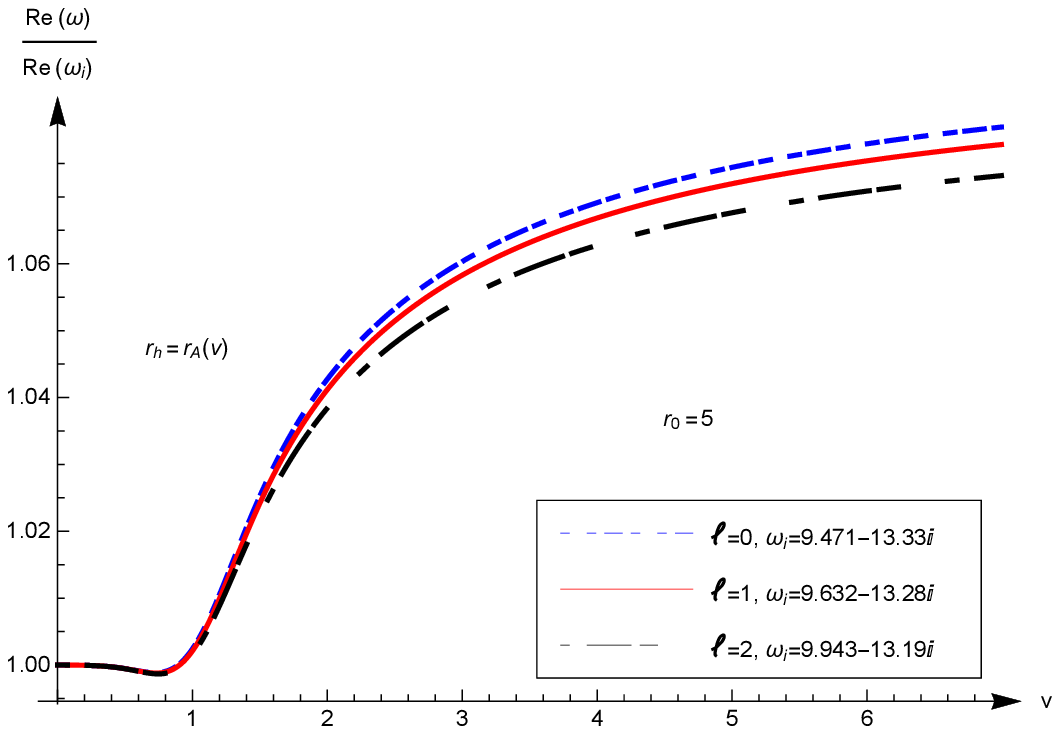}\includegraphics[width=6cm]{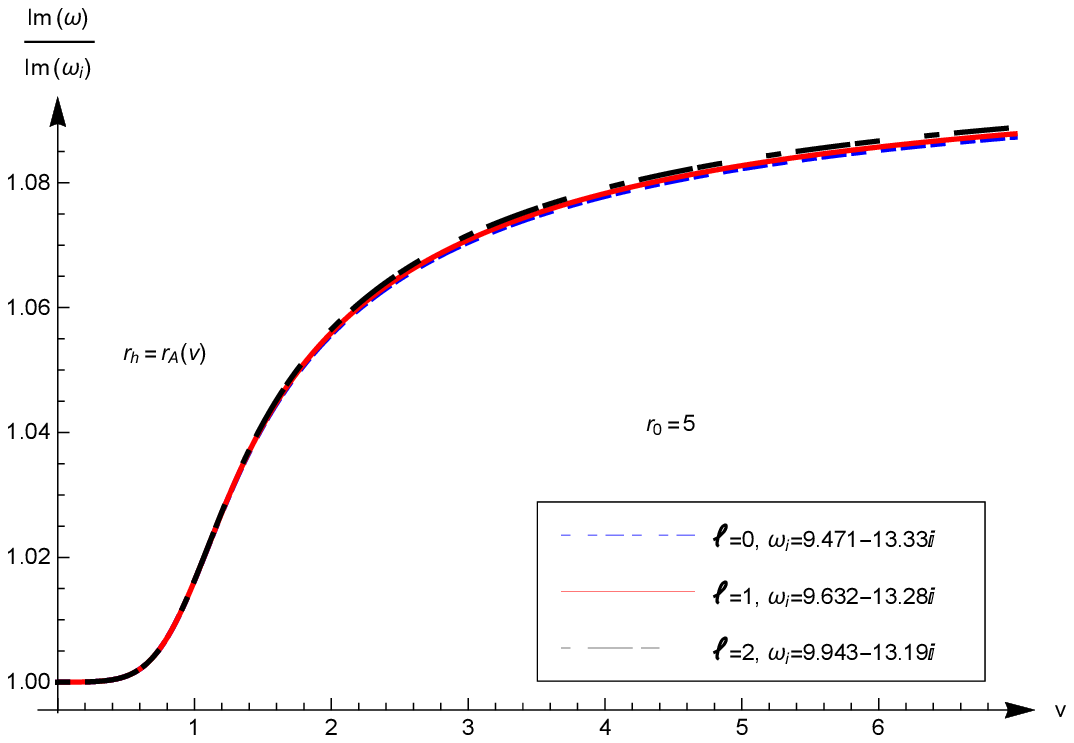}
\includegraphics[width=6cm]{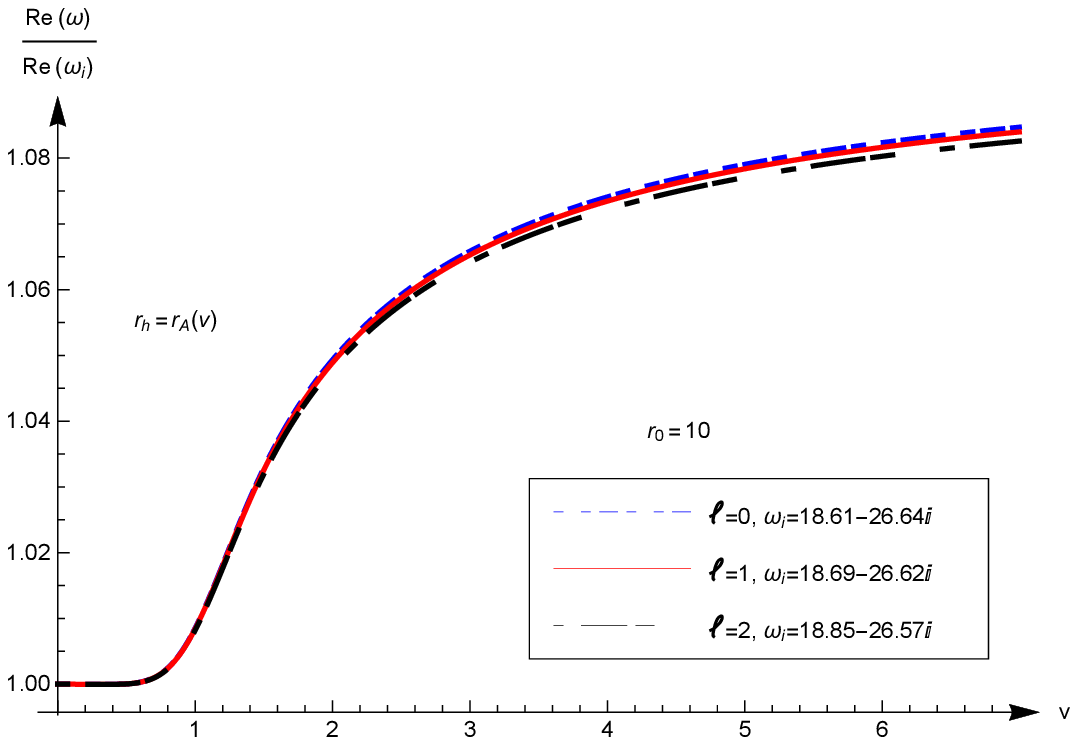}\includegraphics[width=6cm]{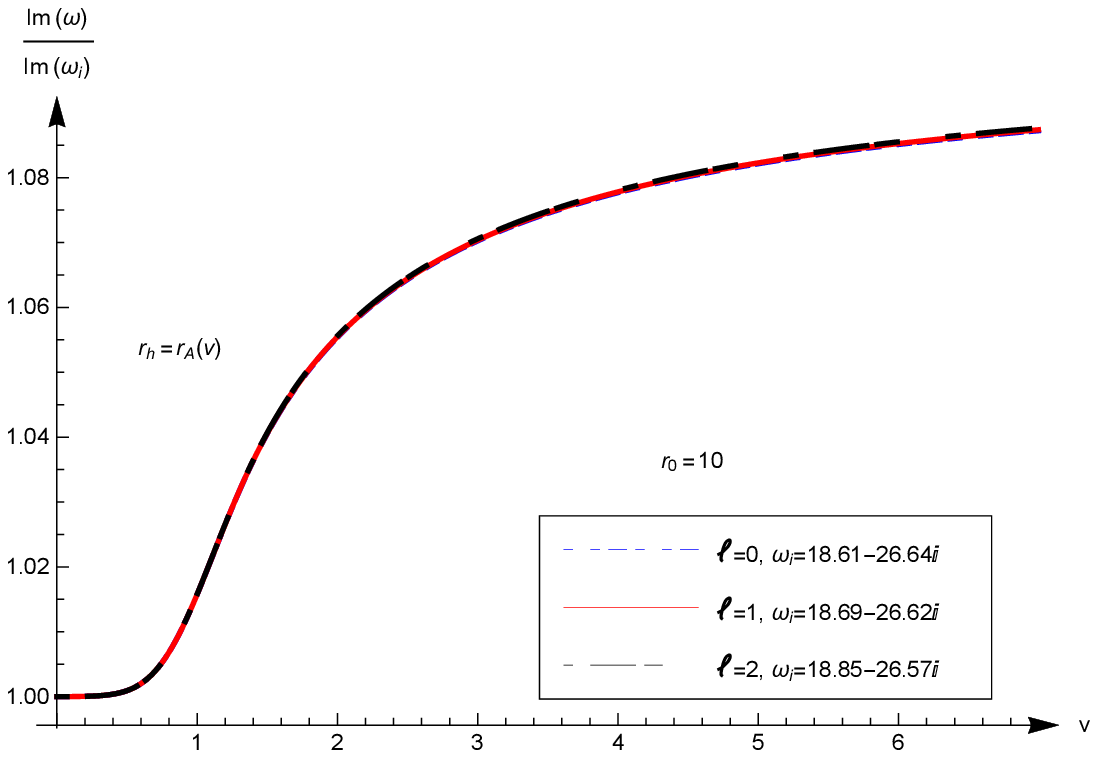}
\caption{(Color online) The real and imaginary parts of the quasinormal frequencies as a function of the Eddington coordinate $v$, where $\omega_i$ is the quasinormal modes frequency associated with the initially static black hole.
The calculations have been carried out for different initial radii $r_0$ as well as angular quantum numbers $\ell$.
The results are presented in terms of the ratios of the quasinormal frequencies to those of static black holes, while the values of the quasinormal frequencies of the corresponding static black holes $\omega_i$ are indicated in the legend.
} \label{QNMf2}
\end{figure*}

\begin{figure*}[h]
\includegraphics[width=6cm]{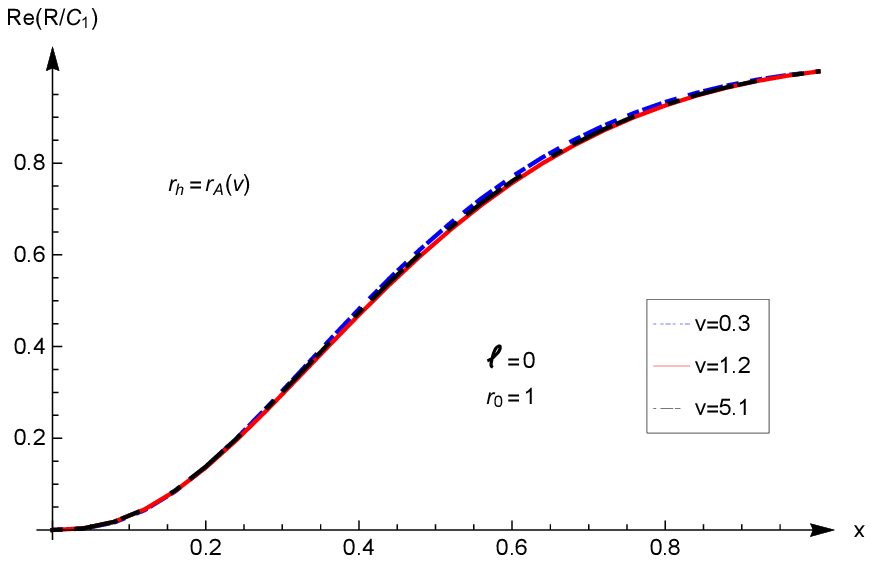}\includegraphics[width=6cm]{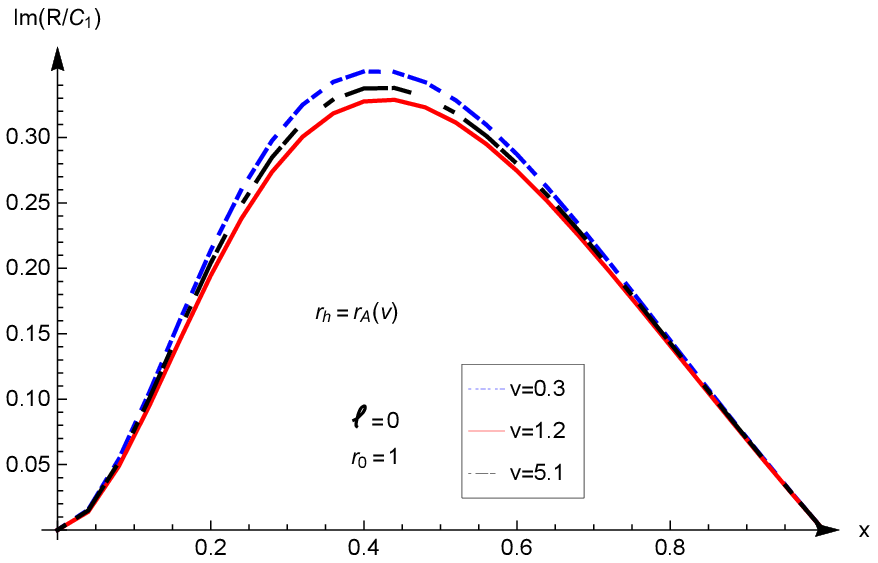}
\caption{(Color online) The radial wave functions $R$ vs. spatial coordinate $x\equiv r_h/r$ of specific quasinormal modes.
Both the real and imaginary parts are normalized by dividing the constant $C_1$ discussed in the text.
The calculations have been carried out for different instants $v$ shown in dashed blue, solid red, and dash-dotted black curves.
} \label{QNMf3}
\end{figure*}

The numerical results are shown in Fig.~\ref{QNMf2} and \ref{QNMf3}.
In Fig.~\ref{QNMf2}, we present the calculated real and imaginary parts of the quasinormal frequencies, for different initially static black holes as well as angular quantum numbers.
Overall, it is found that the quasinormal frequencies of the Vaidya black hole tend to approach those of the corresponding static black holes.
To be specific, as $v \to \infty$, for instance, the obtained quasinormal frequencies approach those of Schwarzschild AdS black holes with $M_2=\frac{r_2^3 + r_2}{2}$.
However, the process takes a more extended period than the duration when the black hole mass evolves from $M_1$ to $M_2$, which numerically terminates at a rather early instant, $v \sim v_2=3/2 $.
In other words, the temporal evolution of the quasinormal frequency exhibit an ``inertial effect", namely, the variation of the quasinormal frequency is delayed in comparison to that of the black hole mass.
This feature has also been observed previously elsewhere~\cite{agr-qnm-time-dependent-03,agr-qnm-time-dependent-04}.

Also, for a given initial value of the apparent horizon, the difference in temporal evolution between different angular quantum numbers increases significantly as the mass of the initially static black hole decreases.
Another nontrivial and interesting feature observed in our calculations is that the real part of the quasinormal frequency does not evolve monotonically before it eventually catches up and approaches the corresponding value of the static black hole.
As shown in the left plot of Fig.~\ref{QNMf2}, instead of immediately following up the value of the corresponding static black hole metric, the real part of the quasinormal frequency decreases first and then increases.
This non-monotonical behavior is found to be less prominent as the initially static black hole becomes more massive.

In Fig.~\ref{QNMf3}, we show the real and imaginary radial parts of the wave functions, evaluated by our numerical scheme.
It is observed that the wave functions thus obtained indeed satisfy the boundary condition discussed in Eq.~(\ref{qnmbc}).
As the wave function is associated with the amplitude of the oscillation, its calculations might turn out to be substantial for future observations.
By employing a more precise numerical scheme proposed in the present study, the above results show once more that the quasinormal modes are, to a first approximation, those of a snapshot of the black hole at the instant when they are computed, corrected by a delay~\cite{agr-qnm-time-dependent-03,agr-qnm-time-dependent-04}.

\section{IV. Discussions and concluding remarks}

In this work, we calculated the quasinormal frequencies of a dynamical black hole background described by Vaidya metric in asymptotically AdS spacetime.
In our calculations, we adopt the apparent horizon to apply the boundary condition.
For a given instant, it is a one-way membrane that only ingoing wave is allowed, and therefore a natural choice for the master equation in question.
The obtained results are reasonable and agree well with the appropriate physical limit of the corresponding static metric.

The introduced scheme is based on a generalized algorithm of the matrix method for the quasinormal modes.
As a result, the proposed approach inherits various advantages of the method.
The resultant quasinormal frequencies are not extracted from the numerical temporal evolution of the perturbations, and therefore, one can achieve a satisfying precision, for both the real and imaginary parts of the frequencies.
Moreover, besides the quasinormal frequencies, the proposed method can be utilized to evaluate the wave function.
Apart from the numerical algorithm itself, in order to generalize the proposed scheme to other dynamical metrics, a vital step of the approach relies on the evaluation of the apparent horizon.
In the specific case of Vaidya metric, the analytic form of the latter is already known.
In a more general context, for example, for the class of metric described by the line element presented in Eq.~(\ref{leDmetric}), the present method can readily be applied, once the apparent horizon coincides with the infinite redshift surface, determined by $g_{vv}=-f(v,r)=0$. 

As the proposed scheme involves the apparent horizon where the boundary condition is exerted, one might be wondering whether the calculated quasinormal frequencies are dependent on the specific choice of coordinate systems.
This seems to be a valid question, as the apparent horizon is defined as the outer component of the intersection of the trapped region and a spacelike surface~\cite{book-blackhole-physics-Frolov}, it depends on the specific coordinate system.
However, if the quasinormal frequencies depend on an arbitrary choice of coordinates, it might potentially undermine the physical content of quasinormal modes for dynamical black hole metrics.
In order to address this issue, let us first fall back to a simpler scenario, the quasinormal modes of a static black hole.
Even for the case, it can be shown that one may also choose a ``non-static" coordinate system, which subsequently modifies the apparent horizon.
Subsequently, both the master equation and its boundary condition are altered.
Obviously, the quasinormal modes of static black hole is a well-defined physical problem, particularly owing to its connection with the results~\cite{adscft-08,adscft-09,adscft-10,hydro-gradex-01} independently obtained via AdS/CFT correspondence~\cite{agr-qnm-holography-review-01}.
Therefore, in this case, one would naturally attribute such apparent ``arbitrariness" regarding the AH solely to the freedom in the choice of the coordinate system.
As a matter of fact, since the black hole is a physical object, it should not rely on the coordinate system describing it.
Similarly, the evolution of small perturbations, namely, the quasinormal modes as a physical process, shall not depend on the choice of coordinates.
Moreover, as it is well-known, quasinormal frequencies are irrelevant to the specific form of the initial perturbations.
In practice, rather than focusing on the vicinity of the black hole horizon, one may consider the spacetime region far away from the black hole horizon.
To be more specific, one may investigate the obtained solution of the master equation by comparing to the asymptotical form at infinity.
As long as the asymptotical properties of spacetime in vacuum are appropriately considered, small perturbations at infinity are well-defined, irrelevant to any specific apparent horizon.
Though the specific numerical value still depends on the coordinates agreed upon between different observers, but it is just a matter of convention, which only concerns the asymptotical properties of the vacuum.
This is because for two observers sitting at infinity who have adopted two distinct coordinate systems, their respective rates of the ``standard clock"s are simply related due to the asymptotically static nature of the spacetime.
By carrying out this procedure, the quasinormal frequencies can be extracted and compared, in the sense that two different observers shall agree with one another.
It is clear that the above argument does not rely on whether the black hole metric is static, and therefore, it can be readily applied to the case of dynamical black holes.
Not surprisingly, for dynamical black holes, the frequency at infinity is in general different from that near the horizon obtained for a given coordinate system~\cite{agr-qnm-lq-matrix-04}.
We plan to apply the proposed method further to other black hole spacetimes in future investigations.

\section*{Acknowledgements}

We gratefully acknowledge the financial support from Brazilian funding agencies Funda\c{c}\~ao de Amparo \`a Pesquisa do Estado de S\~ao Paulo (FAPESP), 
Conselho Nacional de Desenvolvimento Cient\'{\i}fico e Tecnol\'ogico (CNPq), Coordena\c{c}\~ao de Aperfei\c{c}oamento de Pessoal de N\'ivel Superior (CAPES), 
and National Natural Science Foundation of China (NNSFC) under contract No. 11805166.

\section*{Appendix}

\begin{figure*}[h]
\includegraphics[width=6cm]{fig2_Re1.eps}\includegraphics[width=6cm]{fig2_Im1.eps}
\includegraphics[width=6cm]{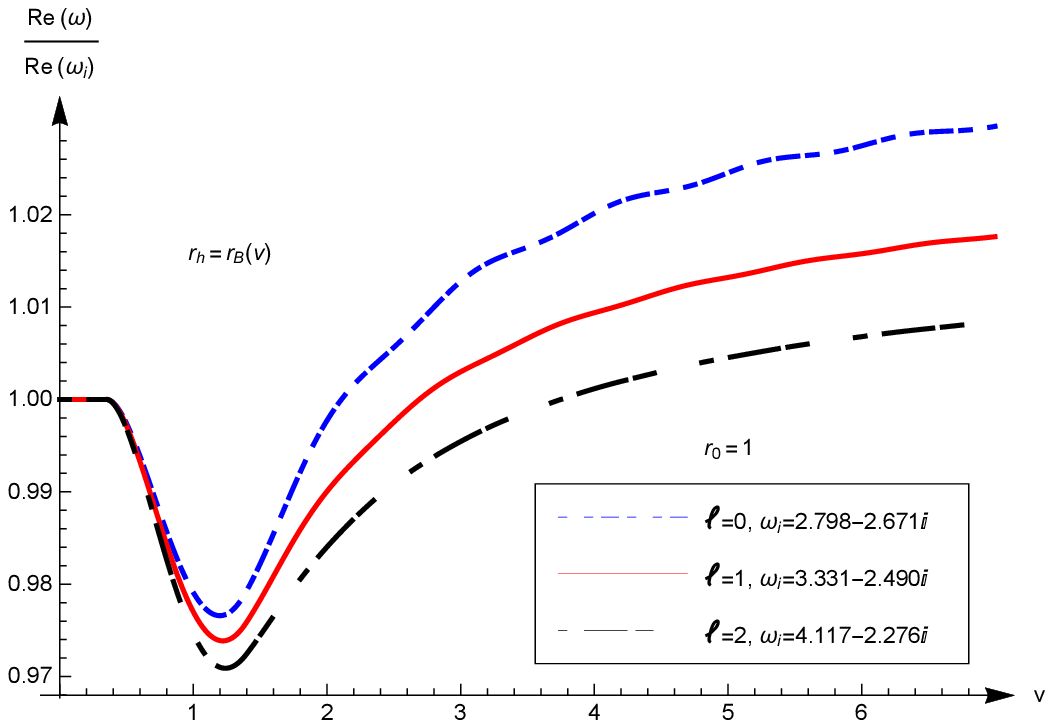}\includegraphics[width=6cm]{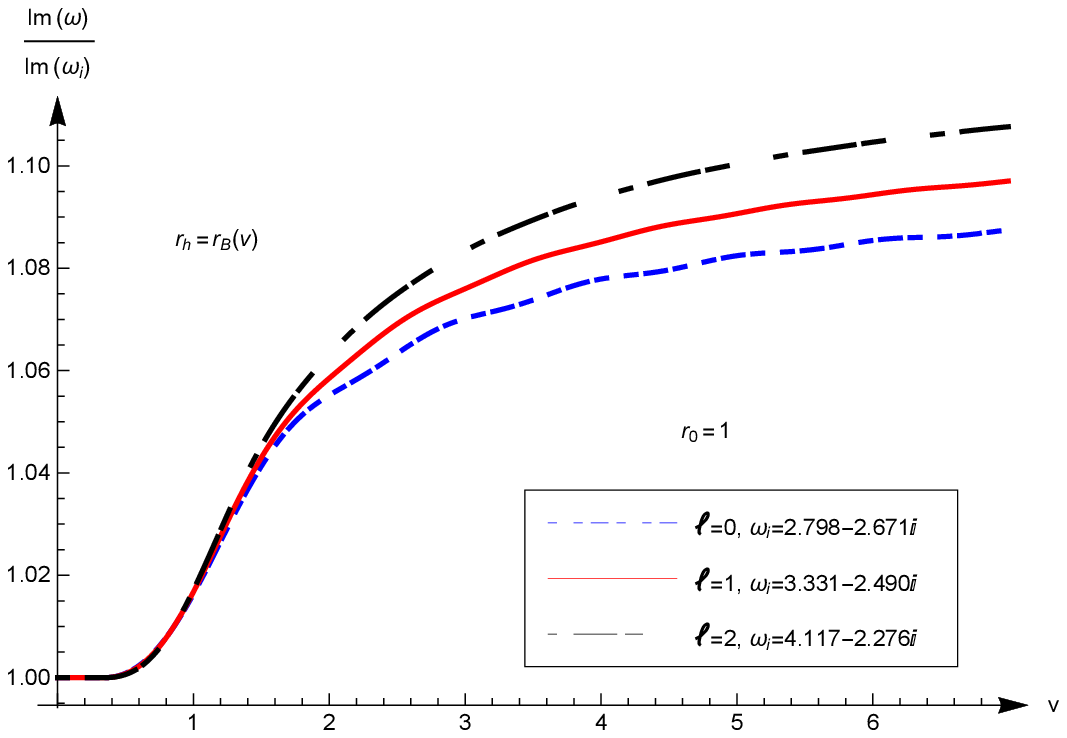}
\includegraphics[width=6cm]{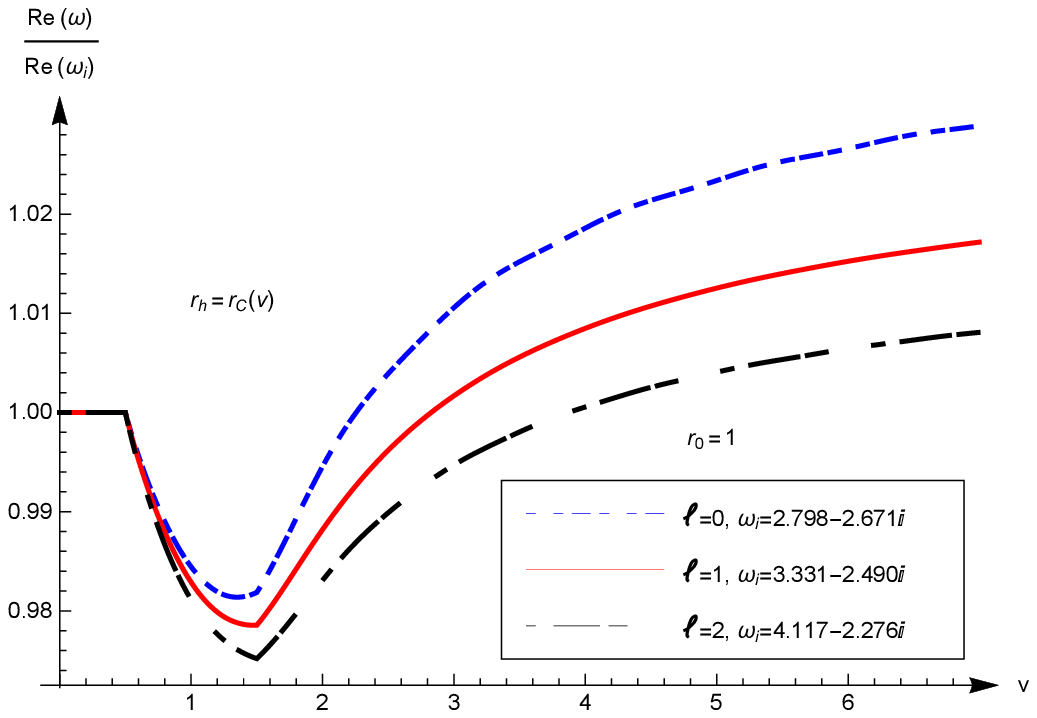}\includegraphics[width=6cm]{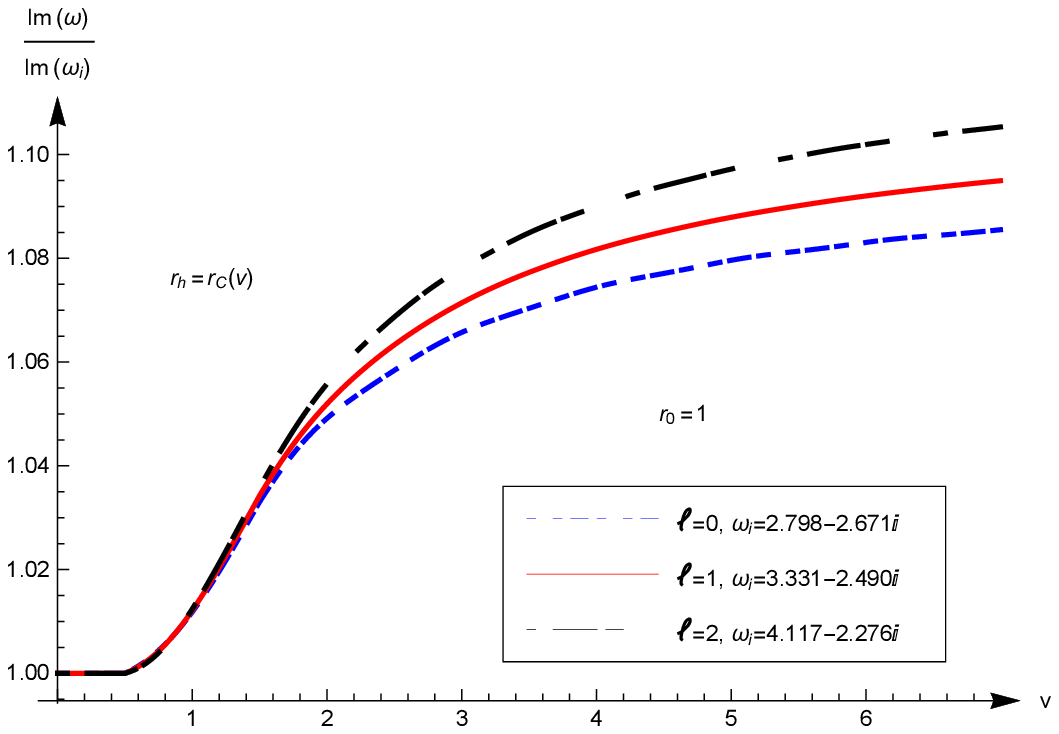}
\caption{(Color online) A comparison between different time-dependent functions $r_A, r_B$ and $r_C$, shown on the top, middle, and bottom rows. 
The real and imaginary parts of the quasinormal frequencies are presented as functions of the Eddington coordinate $v$, where $\omega_i$ is the quasinormal modes frequency associated with the initially static black hole.
The calculations have been carried out for different angular quantum numbers $\ell$, shown in dashed blue, solid red, and dash-dotted black curves respectively.
} \label{QNMf4}
\end{figure*}

In this section, we first show that the results regarding the physical system are manifestly convergent. 
In other words, the obtained numerical results should not be sensitive to small deviations of the chosen mass function.
This is achieved by carrying out the calculations also by two slightly different parameterizations, whose forms have been adopted in some previous studies~\cite{agr-qnm-time-dependent-02,agr-qnm-time-dependent-03,agr-qnm-time-dependent-04}.
These functions are also presented in Fig.~\ref{QNMf1} in solid red and dash-dotted black curves.
\bqn
r_B(v) =
\left\{\begin{array}{cc}
r_1    &  v < v_1^B \cr\\
r_1+\frac{r_2-r_1}{2}\left[1-\cos \left(\frac{v-v_1}{v_2-v_1}\pi\right)\right]  &  v_1^B\le v < v_2^B \cr\\
r_2    &  v_2^B \le v
\end{array}\right. ,\lb{massEx2}
\eqn
where $v_1^B=\frac{7}{20}$, $v_2^B=v_1^B+\frac{\pi}{3}$, and
\bqn
r_C(v) =
\left\{\begin{array}{cc}
r_1    &  v < v_1^C \cr\\
r_1+\frac{v-v_1}{v_2-v_1}(r_2-r_1)  &  v_1^C\le v < v_2^C \cr\\
r_2    &  v_2^C \le v
\end{array}\right. ,\lb{massEx3}
\eqn
where $v_1^C=\frac12$, $v_2^C=v_1^C+1$.
The resultant quasinormal frequencies are presented in Fig.~\ref{QNMf4}.
In particular, we note that even the function $r_C(v)$ is a linear function in $v$, which implies that its first order derivatives are not continuous at $v_1^C$ and $v_2^C$.
The calculated quasinormal frequencies are found to be almost identical despite the small differences between parameterizations.
This, in part, is because the scheme employed in the present study does not explicitly require the derivative to be continuous.
This partly demonstrates that the proposed scheme is indeed reasonably convergent and stable.

Also, we investigate the precision of the present scheme by carrying out calculations using different sizes of the timestep as well as spatial grid size.
The results are shown in Tab.\ref{TableI}.
By using smaller timestep values, it is shown that the numerical results are manifestly convergent.
In particular, the effect of a decrease of two orders of magnitude is found to be insignificant. 
Therefore, the precision of the numerical scheme is admissible for the study of quasinormal modes of dynamical black holes.

\begin{table}[ht]
\caption{\label{TableI} A comparison of the calculated quasinormal frequencies by using different sizes of time interval $\Delta v$ and spatial grid size $\Delta x$.}
\begin{tabular}{cccc}
         \hline
$v$ &~~~~~~$\Delta v=0.02$, $\Delta x=1/19$~~~~~~&~~~~~~$\Delta v=0.001$, $\Delta x=1/25$~~~~~~&~~~~~~$\Delta v=0.0001$, $\Delta x=1/35$~~~~~~\\
        \hline
        \hline
        $0$    &   $2.79827 - 2.67125i$   &     $2.79822 - 2.67121i$  &      $2.79822 - 2.67121i$    \\
        $0.6$  &   $2.78479 -2.67702i$    &     $2.78484 -2.67694i$   &      $2.78492 -2.67692i$     \\
        $1.2$  &   $2.73388 -2.73775i$    &     $2.73393 -2.73763i$   &      $2.73402 -2.73756i$     \\
        $1.8$  &   $2.77454 -2.80484i$    &     $2.77445 -2.8047i$    &      $2.77442 -2.80461i$     \\
        $2.4$  &   $2.80934 -2.83264i$    &     $2.8093 -2.8326i$     &      $2.80931 -2.83259i$     \\
        $3$    &   $2.83292 -2.85662i$    &     $2.83282 -2.85652i$   &      $2.83277 -2.85645i$     \\
        $3.6$  &   $2.8451 -2.86876i$     &     $2.84511 -2.86873i$   &      $2.84515 -2.86872i$     \\
        $4.2$  &   $2.85774 -2.8797i$     &     $2.85762 -2.87965i$   &      $2.85756 -2.87962i$     \\
        $4.8$  &   $2.86317 -2.88725i$    &     $2.86319 -2.88718i$   &      $2.86323 -2.88714i$     \\
        $5.4$  &   $2.87104 -2.89238i$    &     $2.87094 -2.89238i$   &      $2.87088 -2.89239i$     \\
        $6$    &   $2.87444 -2.8983i$     &     $2.87444 -2.89819i$   &      $2.87447 -2.89812i$     \\
        $6.6$  &   $2.87915 -2.90062i$    &     $2.87909 -2.90065i$   &      $2.87907 -2.90069i$     \\
        \hline
\end{tabular}
\end{table}

\bibliographystyle{h-physrev}
\bibliography{references_qian,references_addition}

\end{document}